\documentclass[review]{elsarticle}

\usepackage{lineno,hyperref}
\usepackage{multirow}
\usepackage{array}
\usepackage{booktabs}
\usepackage{arydshln}
\usepackage{amsmath}

\modulolinenumbers[5]

\journal{Optical Materials}

\begin{document}

\begin{frontmatter}

\title{TeO\textsubscript{2}--BaO--Bi\textsubscript{2}O\textsubscript{3} tellurite optical glasses II. - Linear and non-linear optical and magneto-optical properties}

\author[uk,nmsu]{J. Hrabovsky\corref{mycorrespondingauthor}}
\author[nmsu]{J.R. Love}
\author[upce]{L. Strizik}
\author[jap]{T. Ishibashi}
\author[nmsu]{S. Zollner}
\author[uk]{M. Veis}

\cortext[mycorrespondingauthor]{Corresponding author:  jan.hrabovsky@fulbrightmail.org}
\address[uk]{Charles University, Faculty of Mathematics and Physics, Ke Karlovu 3, 121 16 Prague, Czech Republic}
\address[nmsu]{Department of Physics, New Mexico State University, MSC 3D, P.O. Box 30001, Las Cruces, New Mexico 88003-8001, USA}
\address[upce]{University of Pardubice, Faculty of Chemical Technology, Department of General and Inorganic Chemistry,   Studentska 573, Pardubice 532 10, Czech Republic}
\address[jap]{Department of Materials Science and Bioengineering, Nagaoka University of Technology, Kamitomioka 1603-1, Nagaoka 940-2188, Japan}

\begin{abstract}
The present study investigates the linear and non-linear optical and magneto-optical properties of TeO\textsubscript{2}--BaO--Bi\textsubscript{2}O\textsubscript{3} (TeBaBi) glasses prepared by the conventional melt-quenching technique at 900 \textsuperscript{$\circ$}C. Prepared glass composition ranges across the whole glass-forming-ability (GFA) region focusing on mutual substitution trends of constituent oxides, where TeO\textsubscript{2}: 55$-$85 mol.\%, BaO: 10$-$35 mol.\%, Bi\textsubscript{2}O\textsubscript{3}: 5$-$15 mol.\%. Studied glasses exhibit high values of linear ($n$\textsubscript{632} $\approx$ 1.922$-$2.084) and non-linear refractive index ($n$\textsubscript{2}$\approx$1.63$-$3.45$\times10^{-11}$ esu), Verdet constant ($V$\textsubscript{632} $\approx$ 26.7$-$45.3 radT\textsuperscript{$-$1}m\textsuperscript{$-$1}) and optical band gap energy ($E$\textsubscript{g} $\approx$ 3.1$-$3.6 eV). The introduction of TeO\textsubscript{2} and Bi\textsubscript{2}O\textsubscript{3} results in increase of both linear/non-linear refractive index and Verdet constant, with a more pronounced influence of Bi\textsubscript{2}O\textsubscript{3}. Measured spectral dispersion of refractive index and Verdet constant were used for estimation of magneto-optic anomaly parameter ($\gamma \approx$ 0.71--0.92), which may be used for theoretical modelling of magneto-optic response in diamagnetic TeBaBi glasses.  Additionally, the properties of the prepared TeBaBi glasses were directly compared to those of the TeO\textsubscript{2}--ZnO--BaO glass system, which was prepared and characterized under similar experimental conditions. The compositional dependence of the refractive index in both glass systems was described using multilinear regression analysis, demonstrating high correlation and uniformity of estimation across the entire GFA region. This makes them highly promising for precise dispersion engineering and construction of optical devices operating from visible to mid-infrared spectral region.
\end{abstract}

\begin{keyword}
\texttt{} TeO\textsubscript{2}$-$BaO$-$Bi\textsubscript{2}O\textsubscript{3}; tellurite glasses; linear and non-linear optical properties; magneto-optical properties; refractive index; Faraday rotation and Verdet constant
\end{keyword}

\end{frontmatter}

\section{Introduction}
Tellurite glasses, primarily composed of tellurium dioxide, TeO\textsubscript{2}, have become a focal point in optical materials science\cite{mallawany_smart,mallawany_handbook} due to their high linear (pure TeO\textsubscript{2}: $n$\textsubscript{632}=2.178) and non-linear ($n$\textsubscript{2}=3.77$\times$10\textsuperscript{$-$11} esu) refractive index\cite{hrabovsky_pureTeO2} and promising magneto-optical properties represented by Verdet constant ($V$\textsubscript{632}=36.3 radT\textsuperscript{$-$1}m\textsuperscript{$-$1})\cite{hrabovsky_pureTeO2} of approximately one magnitude higher compared to silica-based glasses ($V$\textsubscript{632}=3.7 radT\textsuperscript{$-$1}m\textsuperscript{$-$1})\cite{Weber_1988_verdet_sio2_1,Weber_1988_verdet_sio2_2} without the presence of rare-earth ions. Moreover, wide optical transmission range from visible to mid-infrared (0.4$-$6.5 $\mu$m)\cite{hrabovsky_pureTeO2,HRABOVSKY2024_tebabi_1,HRABOVSKY2024_tzboptika}, low phonon energies\cite{hrabovsky_pureTeO2,Stegeman_2003_ramangain,TAGIARA2017,HRABOVSKY2024_tebabi_1}, good rare-earth ion solubility\cite{mallawany_handbook,Prasad2018,Hrabovsky2025} and notable third-order non-linearity\cite{tebabi_XU2011,Xu2010_nonlin,HRABOVSKY2024_tzboptika} accompanied by low-temperature method of glass synthesis, usually below 1000 \textsuperscript{$\circ$}C\cite{mallawany_handbook}, make them ideal for advanced photonics applications\cite{mallawany_smart}, such as optical fibers and fiber amplifiers\cite{Manning2012,liu_MO,HRABOVSKY2024_tzboptika}, laser host materials\cite{Prasad2018}, Raman amplifiers\cite{Stegeman_2003_ramangain,MANIKANDAN_2017}, etc. Because the synthesis of pure TeO\textsubscript{2} glass using the traditional melt-quenching method is not feasible\cite{TAGIARA2017,hrabovsky_pureTeO2}, tellurite glasses are prepared as multicomponent systems\cite{mallawany_handbook}. This approach not only enhances their glass-forming ability (GFA) but also enables the adjustment of their physical and chemical properties\cite{mallawany_handbook,imaoka1968}. Recent research on the vitrification behavior and structural analysis of TeO\textsubscript{2}--BaO--Bi\textsubscript{2}O\textsubscript{3} (TeBaBi) glasses has extended the GFA region by utilizing compositions with lower TeO\textsubscript{2} content\cite{HRABOVSKY2024_tebabi_1,imaoka1968}. This GFA region enhancement then enables the examination of concentration trends in optical and magneto-optical properties across a wide range of concentrations, as the previous studies on the material properties of ternary TeO\textsubscript{2}--BaO--Bi\textsubscript{2}O\textsubscript{3} glasses were primarily focused on compositions with high TeO\textsubscript{2} content\cite{tebabi_XU2011,tebabi_hill2007,tebabi_Jackson_2009}. Moreover, the previously conducted partial studies contained several mutual differences, such as variations in batch weights, synthesis crucibles/molds, and melting/preheating temperatures (see Supplementary section) \cite{tebabi_XU2011,tebabi_hill2007,tebabi_Jackson_2009} and unlike crystalline materials with defined internal structures, these inconsistencies pose a challenge for accurate comparison. Nevertheless, while the substitution of TeO\textsubscript{2} by other elements typically results in a decrease in refractive index, it has been demonstrated that the addition of Bi\textsubscript{2}O\textsubscript{3} not only partially enhances the glass-forming ability \cite{imaoka1968}, but also increases the refractive index of multicomponent glasses, even when Bi\textsubscript{2}O\textsubscript{3} is substituted for TeO\textsubscript{2} \cite{mallawany_handbook}. This makes TeBaBi glasses highly attractive for photonics applications, highlighting the importance of studying their complex properties across the entire glass-forming ability region. To provide a clearer understanding, Fig. S1 presents a comparison of the TeBaBi samples from this study with those reported in previous research by other authors. 
Based on mentioned findings, the present work extends our previous study\cite{HRABOVSKY2024_tebabi_1} on glass formation, structural and thermal properties of TeBaBi glasses and investigates the linear and non-linear optical and magneto-optical properties of glasses within ternary TeBaBi glass system.

\section{Experimental}
\subsection{Material synthesis and structural characterization}
\begin{figure}[h]
\centering
\includegraphics[width=7cm]{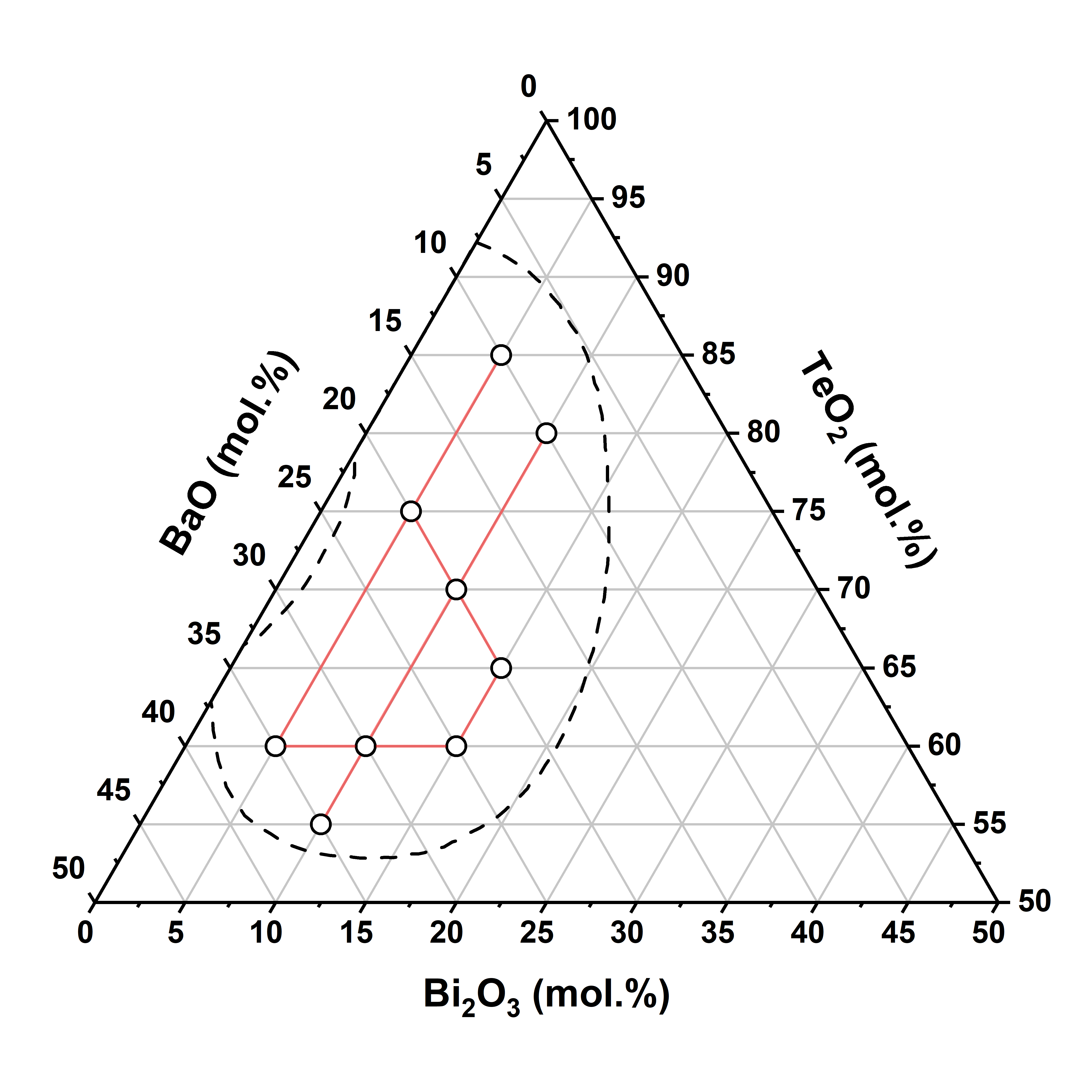}
\caption{\label{fig:ternary_pouzeskla} TeO\textsubscript{2}$-$BaO$-$Bi\textsubscript{2}O\textsubscript{3} (TeBaBi) ternary diagram with indicated glass compositions (black circle) and the boundary of glass-forming region (dashed)\cite{HRABOVSKY2024_tebabi_1,imaoka1968}.}
\end{figure}

The TeO\textsubscript{2}--BaO--Bi\textsubscript{2}O\textsubscript{3} optical glasses were prepared using the conventional melt-quenching technique from high-purity oxides of TeO\textsubscript{2} (Alfa Aesar, 4N), BaO (Sigma Aldrich, 4N) and Bi\textsubscript{2}O\textsubscript{3} (Carl Roth, 5N). The glass-forming ability of TeBaBi glasses together with structural and thermal properties were investigated in our previous work\cite{HRABOVSKY2024_tebabi_1}, whereas this study is focused on nine selected bulk samples across the GFA region and their optical and magneto-optical properties. The investigated samples are listed in Table \ref{tab:tebabi_opt_comp} and shown in Fig.\ref{fig:ternary_pouzeskla} with denoted GFA border. Detailed description of synthesis and materials  shaping/processing conditions, verification of the amorphous state of prepared glasses via X-ray diffraction analysis (XRD) and real chemical composition analyzed by energy-dispersive X-ray (EDX) spectroscopy are provided in Ref.\cite{HRABOVSKY2024_tebabi_1}.

Optical measurements were performed on plane-parallel, one-side polished glassy blocks of thickness $\approx$1.5 mm. The spectral dispersion of optical constants, refractive index ($n$) and extinction coefficient ($k$), was obtained using two variable-angle spectroscopic ellipsometers (VASE) at angles of incidence 65$^{\circ}-$75$^{\circ}$ with 5$^{\circ}$ step in the combined spectral range of 193--30,000 nm. The ellipsometric parameters $\psi$ and $\Delta$ were acquired via RC2 ellipsometer (J.A. Woollam Co., Inc.) in the spectral range of 193--1690 nm (6.42--0.73 eV) with a step of 1 nm and using Fourier-transform IR 
VASE (J.A. Woollam Co., Inc.) from 1700--30 000 nm (5 882--333 cm\textsuperscript{--1}, 50 scans/wavenumber, 15 spectra per revolution with a resolution of  8 cm\textsuperscript{--1}). The optical constants (complex refractive index) of the prepared samples were determined by modeling the combined measured ellipsometric parameters (PSI and DELTA) across the entire spectral range using CompleteEASE software (V6.75b-IR). $\Psi$ and $\Delta$ were fitted to a multi-layer spectroscopic ellipsometry (SE) model composed of semi-infinite bulk material and a layer of surface roughness approximated by Bruggeman effective medium approach\cite{Bruggeman1935,HRABOVSKY2024_tzboptika}.
The magneto-optical (MO) properties of the prepared glasses were investigated using a self-built Faraday rotation setup described in Ref.\cite{Wang2023_mo_jap}, which included a stabilized light source (halogen lamp U-LH100L-3, TH4-100, Olympus Inc.), an electromagnet with a maximum applied field of $\pm1.0$ T, and multi-channel spectral detectors (USB2000+VIS-NIR and NIR Quest 512-2.5, Ocean Optics Inc.) covering a wavelength range of 350--2500 nm. The diamagnetic nature of the prepared glasses was confirmed by measuring Faraday hysteresis loops under varying magnetic fields up to 1 T. For displayed MO measurements shown here, a constant magnetic field of 0.3 T was applied.

\begin{table}[h]
\centering
\caption{Chemical compositions of prepared TeO\textsubscript{2}--BaO--Bi\textsubscript{2}O\textsubscript{3} (TeBaBi) glasses with sample ID and values of density ($\rho$) and molar volume ($V$\textsubscript{M}) taken from Ref.\cite{HRABOVSKY2024_tebabi_1}. The uncertainty of estimation of density and molar volume is $\pm$0.01 g cm\textsuperscript{$-$3} and $\pm$0.05 cm\textsuperscript{3}mol\textsuperscript{$-$1}, respectively.}
\label{tab:tebabi_opt_comp}
\fontsize{8pt}{8pt}\selectfont
\begin{tabular}{cccccc}
\hline
\multirow{2}{*}{SID} & TeO\textsubscript{2}  & BaO  & Bi\textsubscript{2}O\textsubscript{3} & $\rho$  &  $V$\textsubscript{M}     \\
                     & \multicolumn{3}{c}{(mol.$\%$)} &   (g cm\textsuperscript{$-$3})   &    (cm\textsuperscript{3}mol\textsuperscript{$-$1}) \\ \hline
                     Te60Ba25Bi15& 60    & 25   & 15    & 6.16 &  33.11  \\
                     Te65Ba20Bi15& 65    & 20   & 15    & 6.22 &  32.85  \\ \hdashline
                     Te55Ba35Bi10& 55    & 35   & 10    & 5.91 &  31.82  \\
                     Te60Ba30Bi10& 60    & 30   & 10    & 5.94 &  31.71  \\
                              Te70Ba20Bi10 & 70    & 20   & 10    & 6.07 &  31.13  \\
                     Te80Ba10Bi10 & 80    & 10   & 10    & 6.16 &  30.78  \\ \hdashline
                     Te60Ba35Bi5 & 60    & 35   & 5     & 5.7  &  30.30  \\
                             Te75Ba20Bi5& 75    & 20   & 5     & 5.83 &  29.79  \\
                     Te85Ba10Bi5& 85    & 10   & 5     & 6.03 &  28.90  \\ \hline
\end{tabular}
\end{table}

\section{Results and discussion}

The compostion of TeO\textsubscript{2}--BaO--Bi\textsubscript{2}O\textsubscript{3}  glasses was chosen with respect to study the influence of mutual constituent oxides substitutions on linear and non-linear optical and magneto-optical properties. Selected chemical compositions of prepared TeBaBi glasses thus covered entire glass-forming ability region, where TeO\textsubscript{2}: 55$-$85 mol.$\%$, BaO: 10$-$35 mol.$\%$, Bi\textsubscript{2}O\textsubscript{3}: 5$-$15 mol.$\%$.

\begin{figure}[h]
\centering
\includegraphics[width=10.75cm]{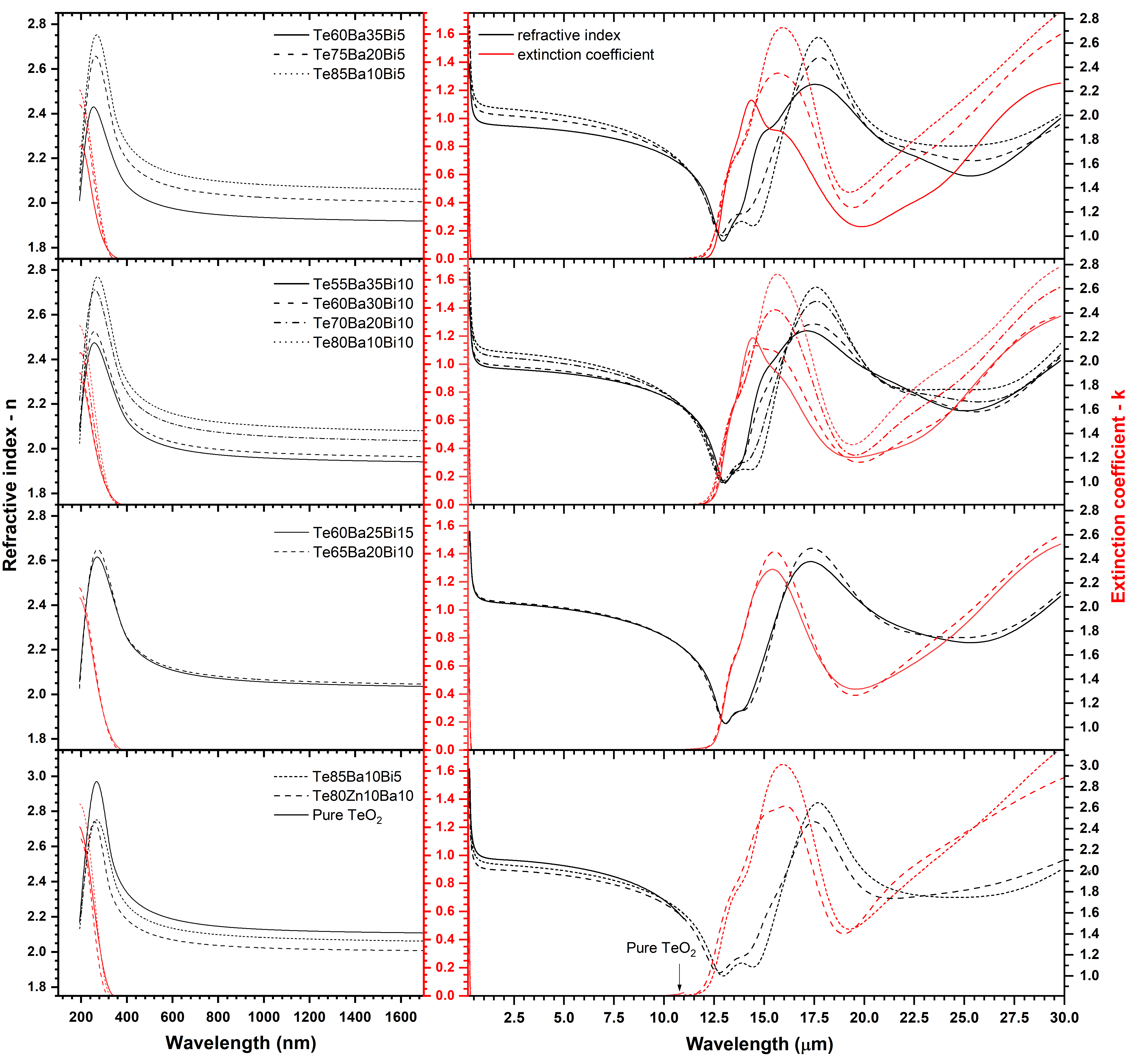}%
\caption{\label{fig:refr_tebabi} Spectral dispersion of refractive index (black lines) and extinction coefficient (red lines) of TeO\textsubscript{2}--BaO--Bi\textsubscript{2}O\textsubscript{3} (TeBaBi) glasses. The bottom part displays the comparison of TeBaBi (this study) and TeO\textsubscript{2}--ZnO--BaO (TZB) glasses from Ref.\cite{HRABOVSKY2024_tzboptika} of approximately equivalent concentration (in at.\%) of the individual constituents with optical constants of pure TeO\textsubscript{2} glass \cite{hrabovsky_pureTeO2}. Note, that spectral dispersion of optical constants for pure TeO\textsubscript{2} covers only the wavelength range 193--11,000 nm.}
\end{figure}

\subsection{Linear optical properties}
\begin{figure}[h]
\centering
\includegraphics[width=8cm]{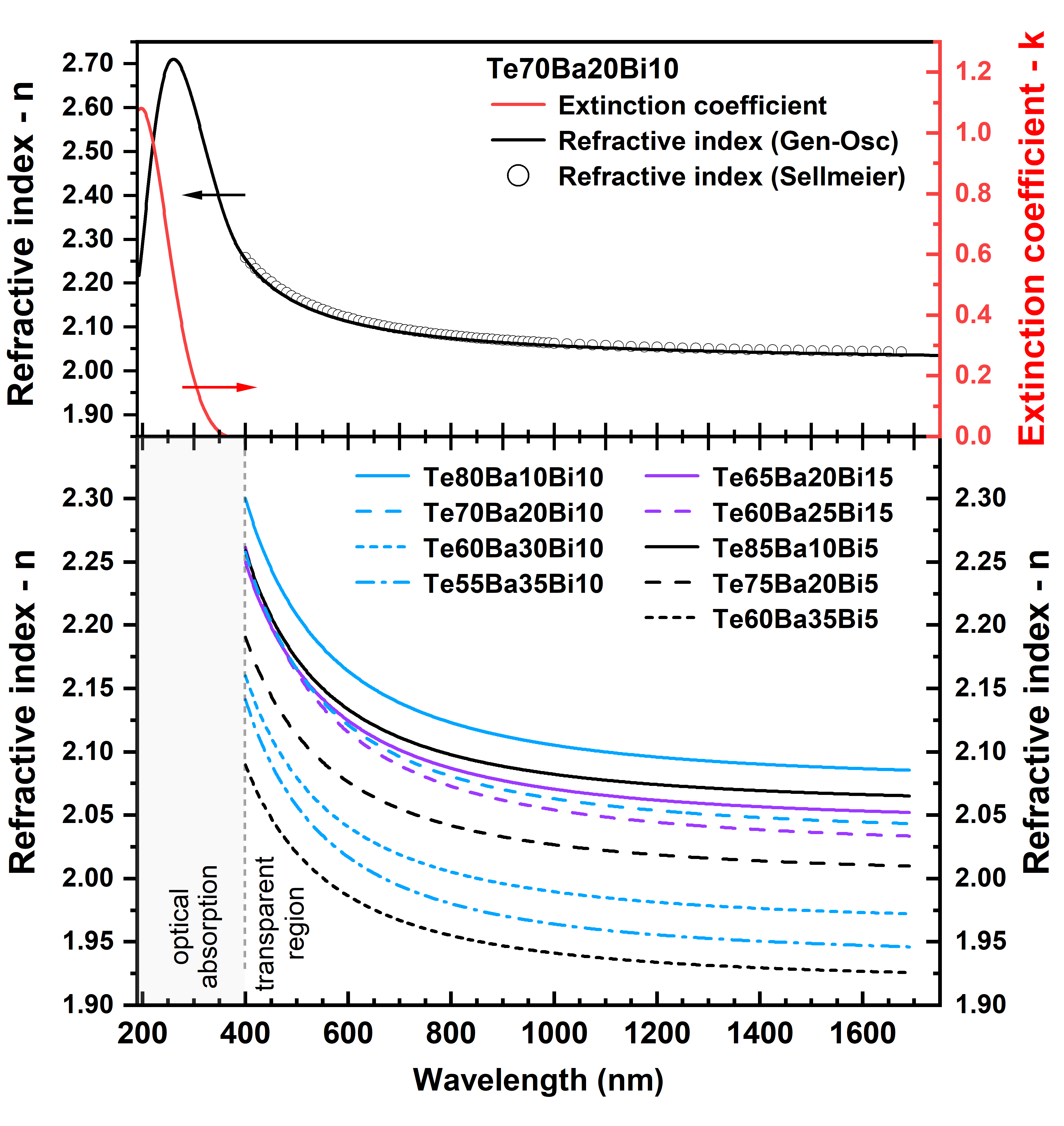}
\caption{\label{fig:tebabi_tern_porov} Comparison between spectral dispersion of refractive index (black line) and extinction coefficient (red line) modeled by Tauc-Lorentz and Sellmeir optical model (open circles) for sample Te70Ba20Bi10 (top part) and spectral dispersion of refractive index of all synthesized TeO\textsubscript{2}--BaO--Bi\textsubscript{2}O\textsubscript{3} (TeBaBi) glasses approximated by Sellmeir model (bottom part).}
\end{figure}

 The spectral dependence of the optical constants $n$ and $k$ for TeBaBi glasses was parametrized using two optical models. The first model coveres the entire experimental spectral range (193--30,000 nm), following the approach in Ref. \cite{HRABOVSKY2024_tzboptika}, and composed of a Tauc-Lorentz (T-L) oscillator for the UV region and a set of Gaussian oscillators for the IR region. The parameters were optimized by minimizing the mean square error between experimental and model data. The extracted optical parameters of TeBaBi glasses are presented in Fig. \ref{fig:refr_tebabi} and Table \ref{tab:tebabi_opt_refr}.  The second model was applied exclusively in the transparent spectral region, below the optical band gap (highlighted in Fig. \ref{fig:tebabi_tern_porov}), extending up to 1700 nm. This model used the standard Sellmeier equation (Eq. \ref{eq:selm_optika_1term}) with two coefficients, $B_1$ and $C_1$ and was later used for theoretical modeling of magneto-optical properties. Both the refractive index and extinction coefficient for all glasses are enclosed in the Supplementary information in the whole observed spectral range.

\begin{equation}
\label{eq:selm_optika_1term}
n\textsuperscript{2} = 1+ \frac{ B\textsubscript{1}\lambda\textsuperscript{2}}{\lambda\textsuperscript{2}-C\textsubscript{1}} 
\end{equation}

The consistency between both optical models is demonstrated in Fig.~\ref{fig:tebabi_tern_porov} (top), using the Te70Ba20Bi10 sample as a representative case, where an agreement between the fitted curves confirms their compatibility across the relevant spectral range. The extracted refractive index values, $n_{\lambda}$, at wavelengths  $\lambda$ = 632  nm and  $\lambda$ = 1550  nm are presented in Table \ref{tab:tebabi_opt_refr}. Additionally, the table includes the molar refractivity, ($R_M$) and electronic polarizability ($\alpha$) derived using Eq. \ref{eq:selm_optika_refractivity} and Eq. \ref{eq:selm_optika_polarizability}, respectively, where $N_A$ represents Avogadro’s number. For the calculation of $R$\textsubscript{M} and $\alpha$, the refractive index value at 1550 nm, corresponding to the spectral region with minimal dispersion, was employed.

\begin{equation}
\label{eq:selm_optika_refractivity}
R\textsubscript{M} = \left( \frac{n\textsuperscript{2}-1}{n\textsuperscript{2}+2}\right) \times V\textsubscript{M}
\end{equation}

\begin{equation}
\label{eq:selm_optika_polarizability}
\alpha = \left( \frac{3}{4 \pi N\textsubscript{A}}\right) \times R\textsubscript{M}
\end{equation}

\begin{table}[]
\centering
\caption{Optical parameters of TeO\textsubscript{2}$–$BaO$-$Bi\textsubscript{2}O\textsubscript{3} (TeBaBi) glasses: refractive index ($n$\textsubscript{$\lambda$}) at wavelengths of $\lambda \approx$ 632 nm or 1550 nm, Abbe number ($\nu$\textsubscript{D}), molar refractivity ($R$\textsubscript{M}), electronic polarizability ($\alpha$) and Sellmeier parameters for description of spectral dependence of refractive index ($B$\textsubscript{1}, $C$\textsubscript{1}) according to Eq.\ref{eq:selm_optika_1term}. The refractive index and molar refractivity error range is $\pm$0.001 and $\pm$0.01 cm\textsuperscript{3}mol\textsuperscript{$-$1}, respectively.  Numbers in parentheses represent the tolerance on the last digit.}
\label{tab:tebabi_opt_refr}
\fontsize{7pt}{7pt}\selectfont
\begin{tabular}{cccccccc}
\hline
\multirow{2}{*}{SID} & $n$\textsubscript{632}  & $n$\textsubscript{1550} & $\nu$\textsubscript{D}   &$R$\textsubscript{M,1550}     & $\alpha$ & B1      & C1 \\
                     &       &       &      &(cm\textsuperscript{3}mol\textsuperscript{$-$1})&  (\AA\textsuperscript{3})     &         &     ($\times$10\textsuperscript{$-$3} $\mu$m\textsuperscript{2})      \\ \hline
                     Te60Ba25Bi15& 2.100
& 2.038
& 17.0
&16.97
& 6.73
& 3.09(1) & 39.7(2)     \\
                     Te65Ba20Bi15& 2.109& 2.049& 16.7
&16.94& 6.72& 3.17(1) & 35.0(1)     \\ \hdashline
                     Te55Ba35Bi10& 1.995
& 1.937
& 17.4
&15.23
& 6.04
& 2.75(1) & 37.2(2)     \\
                     Te60Ba30Bi10& 2.023
& 1.967
& 18.0
&15.50
& 6.14
& 2.85(1)  & 35.4(1)     \\
                              Te70Ba20Bi10 & 2.103& 2.039& 17.5
&15.96& 6.33& 3.13(1) & 37.6(1)     \\
                     Te80Ba10Bi10 & 2.149
& 2.084
& 17.0
&16.22
& 6.43
& 3.31(1) & 36.7(1)     \\ \hdashline
                     Te60Ba35Bi5 & 1.970
& 1.922
& 20.0
&14.33
& 5.68
& 2.68(1) & 32.7(1)    \\
                             Te75Ba20Bi5& 2.066
& 2.008
& 18.4
&14.98
& 5.94
& 3.00(1) & 33.4(1)     \\
                     Te85Ba10Bi5& 2.126& 2.064& 18.0&15.06& 5.97& 3.22(1) & 33.7(1)     \\ \hline
\end{tabular}
\end{table}

\begin{figure}[]
\centering
\includegraphics[width=13cm]{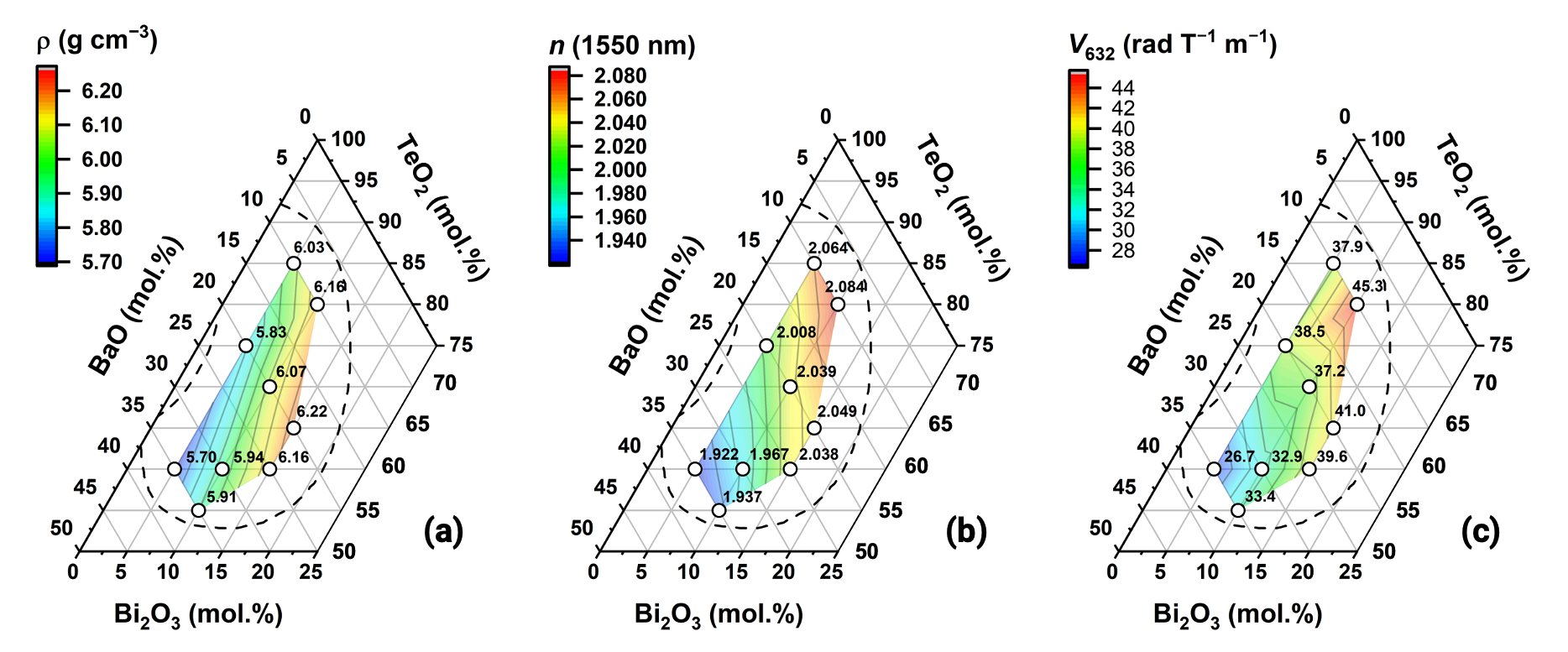}
\caption{\label{fig:tebabi_tern_rho_n_v} Evolution of glass density, $\rho$, (taken from Ref.\cite{HRABOVSKY2024_tebabi_1}) refractive index, $n_{1550}$, and
Verdet constant, $V_{632}$, with respect to chemical composition of TeO\textsubscript{2}--BaO--Bi\textsubscript{2}O\textsubscript{3} (TeBaBi) glass.}
\end{figure}

The refractive index values at 1550 nm vary from 1.922 to 2.084 in the GFA region, as shown in Fig.\ref{fig:tebabi_tern_rho_n_v}. The obtained refractive index values are then slightly lower than for pure TeO\textsubscript{2} glass ($n$\textsubscript{1550}=2.111)\cite{hrabovsky_pureTeO2}, but higher compared to the other commonly used ternary (TeO\textsubscript{2}$-$ZnO$-$Na\textsubscript{2}O, TeO\textsubscript{2}$-$ZnO$-$BaO) or binary (TeO\textsubscript{2}$-$BaO) glasses\cite{mallawany_handbook,HRABOVSKY2024_tzboptika}. The compositional evolution of refractive index for TeBaBi glasses may be described using the multilinear regression analysis (Eq.\ref{eq:tebabi_multilin_refractive_mol})

\begin{equation}
\label{eq:tebabi_multilin_refractive_mol}
n_\text{1550}= a\textsubscript{TeO2} \cdot m(\text{TeO}\textsubscript{2}) + a\textsubscript{BaO} \cdot m\text{(BaO)}+a\textsubscript{Bi2O3} \cdot m(\text{Bi}\textsubscript{2}\text{O}\textsubscript{3}),
\end{equation}

where $a$\textsubscript{X} are the multilinear fit parameters ($a$\textsubscript{TeO2}=21.0$\times$10\textsuperscript{$-$3}, $a$\textsubscript{BaO}=15.2$\times$10\textsuperscript{$-$3}, $a$\textsubscript{Bi2O3}= 25.9$\times$10\textsuperscript{$-$3}) and $m$(X) represents the molar content (in mol.$\%$) of constituent oxides TeO\textsubscript{2}, BaO and Bi\textsubscript{2}O\textsubscript{3}. Estimated correlation coefficient was $r\geq 0.993$ with the maximum observed deviation ($\Delta n$\textsubscript{M}) value between the predicted and experimentally determined refractive index of about $\approx0.01$. Multilinear optimization can be also conducted with respect to the atomic percent content (Eq.\ref{eq:tebabi_multilin_refractive_ato}) of the three respected cations, given the similarly good correlations $r\geq 0.994$, with corresponding multilinear parameters: $a$\textsubscript{Te}=62.9$\times$10\textsuperscript{$-$3}, $a$\textsubscript{Ba}=25.9$\times$10\textsuperscript{$-$3} and $a$\textsubscript{Bi}= 58.7$\times$10\textsuperscript{$-$3}

\begin{equation}
\label{eq:tebabi_multilin_refractive_ato}
n_\text{1550}= a\textsubscript{Te} \cdot m(\text{Te}) + a\textsubscript{Ba} \cdot m\text{(Ba)}+a\textsubscript{Bi} \cdot m(\text{Bi}).
\end{equation}

Given the strong correlation and uniformity in refractive index estimation across the entire glass-forming ability (GFA) region, the presented results pave the way for the controlled synthesis of TeBaBi glasses with tailored refractive indices, thereby enabling precise refractive index engineering in the transparent, low-dispersion spectral region. A similar conclusion was reached for the TeO\textsubscript{2}--ZnO--BaO glass system, which was synthesized and measured under the similar conditions\cite{HRABOVSKY2022_tzb,HRABOVSKY2024_tzboptika}. The following multilinear fit coefficients define the relationship between $n$\textsubscript{1550} and the molar contents of the constituent TZB oxides (data akin to those from Eq.\ref{eq:tebabi_multilin_refractive_mol} and Eq.\ref{eq:tebabi_multilin_refractive_ato}) $a$\textsubscript{TeO2}=21.4$\times$10\textsuperscript{$-$3}, $a$\textsubscript{ZnO}=17.0$\times$10\textsuperscript{$-$3} $a$\textsubscript{BaO}=14.8$\times$10\textsuperscript{$-$3} ($r$=0.995; $\Delta n$\textsubscript{M}$\approx$0.01) or atomic cation's percentage $a$\textsubscript{Te}=63.5$\times$10\textsuperscript{$-$3}, $a$\textsubscript{Zn}=32.4$\times$10\textsuperscript{$-$3} $a$\textsubscript{Ba}=26.6$\times$10\textsuperscript{$-$3} ($r$=0.995; $\Delta n$\textsubscript{M}$\approx$0.01). An important observation arising from the comparison of both systems is that the multilinear coefficients associated with the common components, TeO\textsubscript{2} and BaO (or their atomic representations), remain nearly identical regardless of the selected ternary system.

Within the TeBaBi glass system, the refractive index is further observed to increase with both the substitution of TeO\textsubscript{2} for BaO at constant Bi\textsubscript{2}O\textsubscript{3} content and with the addition of Bi\textsubscript{2}O\textsubscript{3}, irrespective of the other constituents. From Fig.\ref{fig:tebabi_tern_rho_n_v},  it is evident that the observed compositional refractive index behavior can be linked to the evolution of glass density. This behavior may also be explained by changes in molar refractivity and/or electronic polarizability (see Table \ref{tab:tebabi_opt_refr}). Refractive index changes were also previously associated with the decay of glass structure and transformation of [TeO\textsubscript{4}] to [TeO\textsubscript{3}] structural units which is connected with a change in the optical basicity\cite{HRABOVSKY2022_tzb}. Based on this assumption, the Te85Ba10Bi5 glass should exhibit a refractive index value similar to that of the TeO\textsubscript{2}$-$ZnO$-$BaO  glass (Te80Zn10Ba10) with a comparable composition (in at.$\%$) due to the observed similar degree of internal glass structure transformation derived from Raman scattering measurements\cite{HRABOVSKY2024_tebabi_1}. However, the observed refractive index difference between these two glasses is approximately $\Delta n \approx$0.05, which is significantly larger value than the experimental measurement uncertainty. It is therefore evident that the absolute value of the refractive index is influenced not only by the degree of internal structure transformation but mainly by the presence of particular constituent oxides/atoms in the glass. Obtained results for several overlapping glass compositions were compared with values of refractive index at 1550 nm (obtained by prism-coupled refractometry method) presented in the work of Hill et al.\cite{tebabi_hill2007}, where the observed deviation was within $\pm$0.01. Further note that the discussed difference is comparable to the prediction accuracy of the above-mentioned multilinear fits. The results presented here are, thus, in good agreement and demonstrate satisfactory material reproducibility and predictability within the TeBaBi glass system. 

The degree of light dispersion was investigated by estimation of the Abbe number, using the known values of linear refractive index at wavelengths of the Fraunhofer C, D and F spectral lines (656.3 nm, 587.6 and 486.1 nm) as follows: $\nu\textsubscript{D}= (n\textsubscript{587.6}-1)/(n\textsubscript{486.1}-n\textsubscript{656.3})$. Obtained Abbe numbers of TeBaBi glasses were in the range 16.7$-$20.0 which is comparable to the value of pure TeO\textsubscript{2} glass (19.9),\cite{hrabovsky_pureTeO2} or other TeO\textsubscript{2}-rich multicomponent tellurite glasses (18.5$-$19.1)\cite{mallawany_handbook,HRABOVSKY2024_tzboptika}, but slightly lower compared to TZB glasses (18.5$-$26.3) prepared under similar synthesis conditions\cite{HRABOVSKY2024_tzboptika}. This can be connected with the position of short-wavelength absorption edge, which is situated more in UV part of spectra for TZB glasses compared to TeBaBi glass system\cite{HRABOVSKY2024_tzboptika,HRABOVSKY2024_tebabi_1}. Pure TeO\textsubscript{2} glass as well as the majority of multicomponent tellurite glasses thus belong to the group of very dense flint glasses.

The optical band gap energy ($E_g$) of the studied glasses was determined using the Tauc–Lorentz (T–L) oscillator model, following the methodology outlined in Refs. \cite{HRABOVSKY2024_tzboptika,hrabovsky_pureTeO2}. The evaluated $E_g$ values fall within the range of approximately 3.09 to 3.57 eV (Table \ref{tab:tebabi_nonlinear}) and show a clear increasing trend with higher TeO\textsubscript{2} and BaO content. Observed increase in $E_g$ while substituting Bi atoms by Ba can be explained on the basis of decreasing refractive index according to the Lorentz model (Eq. \ref{eq:lor}), where $\epsilon (\infty)$ is the high-frequency permittivity,  $\omega_p$ and $\omega_0$ are the plasma frequency and the resonance frequency of the oscillator. Applying the $\omega \rightarrow 0$ approximation to Eq.~\ref{eq:lor}, the refractive index becomes inversely proportional to the spectral position of the used oscillator, and thus also to the optical band gap energy. This relationship is commonly referred to as the Moss rule. However, in the present case, the substitution of Te for Ba at constant Bi content results in a simultaneous increase in both the refractive index and the optical band gap energy. The employed Lorentz model approximation further indicates that the refractive index is not solely governed by the inverse dependence on the band gap energy, but is also directly proportional to the plasma frequency and, consequently, to the electron density. The replacement of barium with tellurium increases the electron density of the material, which likely exerts a more significant influence on the observed variation in band gap energy. 

\begin{equation}
\label{eq:lor}
\epsilon (\omega) = \epsilon (\infty)+\frac{\omega_p^2}{\omega_0^2-\omega^2-i\omega\gamma_{lor}} \rightarrow \epsilon (\omega) = \epsilon (\infty)+\frac{\omega_p^2}{\omega_0^2} \text{   for } \omega \rightarrow0 
\end{equation}

In general, obtained $E_g$ values for TeBaBi glasses are lower than that of pure TeO\textsubscript{2} glass ($E_g$ = 3.72 eV)\cite{hrabovsky_pureTeO2}. In comparison to the reference TZB glass system\cite{HRABOVSKY2022_tzb,HRABOVSKY2024_tzboptika}, it is evident that the zinc-containing glasses exhibit higher values of optical band gap energy (T80Zn10Ba10: $E_g=3.86 $ eV vs T85Ba10Bi5: $E_g=3.57$ eV), whereas posses a lower refractive index T80Zn10Ba10: $n_{1550}=2.009$ eV vs T85Ba10Bi5: $n_{1550}=2.064$ eV), which is in accordance with aforementioned Moss rule.

\begin{table}[]
\caption{Optical parameters of TeO\textsubscript{2}--BaO--Bi\textsubscript{2}O\textsubscript{3} (TeBaBi) glasses: optical band gap energy ($E$\textsubscript{g}), single oscillator ($E_0$) and dispersion ($E_d$) energy, non-linear refractive index ($n_2$) and refractive index for photon energies $E \rightarrow 0$ eV ($n$(0)), and first-order ($\chi$\textsuperscript{(1)}) and third-order ($\chi$\textsuperscript{(1)}) optical susceptibility. The estimation error of optical band gap energy, dispersion energy and single oscillator energy is approximately 0.02 eV. The refractive index $n$(0) and non-linear refractive index $n_2$ error range is $\pm$0.01 }
\label{tab:tebabi_nonlinear}
\fontsize{7pt}{7pt}\selectfont
\begin{tabular}{lccccccc}
\hline
SID            & $E$\textsubscript{g}    & $E$\textsubscript{0}   & $E$\textsubscript{d}    & $n$(0) & $\chi$\textsuperscript{(1)}     & $\chi$\textsuperscript{(3)}     & $n$\textsubscript{2}       \\
               & \multicolumn{3}{c}{(eV)} &      & (esu) & ($\times$10\textsuperscript{$-$13} esu) &  ($\times$10\textsuperscript{$-$11} esu) \\ \hline
Te60Ba25Bi15   & 3.16& 6.72 & 20.98 & 2.03 & 0.25     & 6.47     & 2.76     \\
Te65Ba20Bi15   & 3.29& 6.76 & 21.49 & 2.04 & 0.25     & 6.95     & 2.95     \\ \hdashline
Te55Ba35Bi10   & 3.09& 6.99 & 19.43 & 1.94 & 0.22     & 4.06     & 1.81     \\
Te60Ba30Bi10   & 3.20  & 7.01 & 20.13 & 1.97 & 0.23     & 4.65     & 2.05     \\
Te70Ba20Bi10   & 3.33& 6.79 & 21.43 & 2.04 & 0.25     & 6.75     & 2.87     \\
Te80Ba10Bi10   & 3.39& 6.72 & 22.30 & 2.08 & 0.26     & 8.27     & 3.45     \\ \hdashline
Te60Ba35Bi5    & 3.25& 7.41 & 20.00 & 1.92 & 0.21     & 3.62     & 1.63     \\ 
Te75Ba20Bi5    & 3.54& 6.88 & 20.60 & 2.00 & 0.24     & 5.49     & 2.38     \\
Te85Ba10Bi5    & 3.57& 6.83 & 22.02 & 2.06 & 0.26     & 7.38     & 3.11     \\ \hline
T80Zn10Ba10\cite{HRABOVSKY2024_tzboptika}    & 3.86  & 7.00    & 20.96 & 2.00    & 0.238    & 5.49     & 2.38     \\ 
Pure TeO\textsubscript{2} (Pt)\cite{hrabovsky_pureTeO2} &   3.72    & 6.47 & 22.00    & 2.10  & 0.27     & 9.22     & 3.77     \\
g-SiO\textsubscript{2}\cite{wemple_1973_WDD_sio2}         &       & 13.4 & 14.7  & 1.45 & 0.087    & 0.0987   & 0.0591   \\
KH\textsubscript{2}PO\textsubscript{4}\cite{wemple_1971_WDD_ostatni}         &       & 12.8 & 16    & 1.5  & 0.099    & 0.166    & 0.0962   \\
LiNbO\textsubscript{3}\cite{wemple_1971_WDD_ostatni}         &       & 6.65 & 25.9  & 2.21 & 0.31     & 15.7     & 6.15     \\ \hline
\end{tabular}
\end{table}

\subsection{Non-linear optical properties}
Spectral dependence of refractive index in the transparent optical region has been further used similarly as in Ref.\cite{HRABOVSKY2024_tzboptika} for the determination of dispersion energy ($E\textsubscript{d}$) and single oscillator energy ($E\textsubscript{0}$) utilizing the single-oscillator approximation according to Wemple and DiDomenico (WDD)\cite{wemple_1,wemple_2}. The intercept and slope of the linear fit to $(n^2 – 1)\textsuperscript{-1}$ versus square of photon energy $E^2$ dependency in transparent spectral region provides parameters $E\textsubscript{d}$ and $E\textsubscript{0}$ ( Eq.\ref{eq:wdd_1}), which are used for estimation of refractive index $n_0$ as $n_0 = \left[ 1+(E_{d,W}/E_{0,W})\right]^{1/2}$.

\begin{equation}
\label{eq:wdd_1}
n^2-1= \frac{E_{0}E_{d}}{E_{0}^2-E^2}
\end{equation}

The non-linear refractive index, $n_2$, for TeBaBi glasses has been determined by the procedure described by Ticha \& Tichy\cite{ticha_2002}. First, the linear optical susceptibility ($\chi$\textsuperscript{(1)}) was calculated using  $\chi$\textsuperscript{(1)} = $(n^2$-- 1)/4$\pi$ at photon energy $E\rightarrow 0$ by using Eq.\ref{eq:wdd_1} resulting in Eq.\ref{eq:wdd_2}(a). Based on that, the third-order optical susceptibility $\chi$\textsuperscript{(3)} was calculated by Eq.\ref{eq:wdd_2}(b).

\begin{equation}
\label{eq:wdd_2}
\text{(a) }\chi^{(1)}= \frac{1}{4\pi}\frac{E_{d,W}}{E_{0,W}}  ;  \text{(b) }\chi^{(3)}= A_{W}\left[ \chi^{(1)}\right]^4
\end{equation}

where \textit{A}\textsubscript{W} = 1.7$\times$10\textsuperscript{--10} for $\chi$\textsuperscript{(1)} in esu units \cite{ticha_2002}. Subsequently, the non-linear refractive index was estimated by using Eq.\ref{eq:wdd_4} including the correction recommended by P.Gorski et al.\cite{Gorski1996_wdd}. Derived values are listed in Table \ref{tab:tebabi_nonlinear}.

\begin{equation}
\label{eq:wdd_4}
n_2\approx2.3\frac{12\pi\chi^{(3)}}{n_0}
\end{equation}

The non-linear refractive index of studied TeBaBi glasses is in range from 1.63$\times$10\textsuperscript{--11} esu to 3.45$\times$10\textsuperscript{--11}
esu and follows similar concentrations trends as the linear refractive index. The highest value was therefore obtained for Te80Ba10Bi10 composition which is
lower compared to pure TeO\textsubscript{2} glass ($n_2$ = 3.77 $\times$10\textsuperscript{--11} esu)\cite{hrabovsky_pureTeO2}. Mutual comparison of previously discussed close compositions from TeBaBi and TZB
glass systems revealed the positive difference of $\Delta n_2 \approx$0.7$\times$10\textsuperscript{--11} esu\cite{HRABOVSKY2024_tzboptika}. It can be therefore concluded, that both ternary systems of TZB and TeBaBi glass systems are promising for non-linear optical applications as can be also seen via comparison with other commonly used materials listed in Table \ref{tab:tebabi_nonlinear}.

\begin{figure}[h]
\centering
\includegraphics[width=13cm]{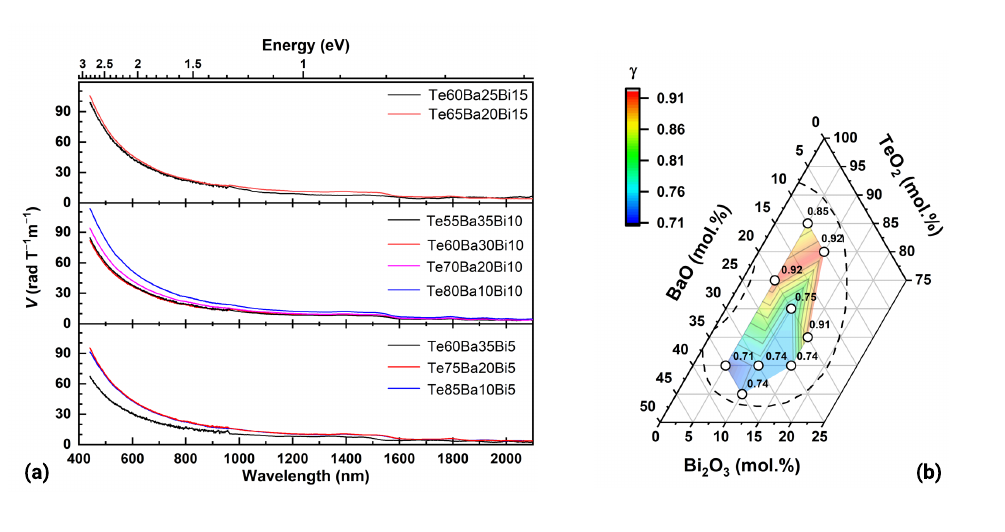}
\caption{\label{fig:mo} (a) Spectral dispersion of Verdet constant of TeO\textsubscript{2}--BaO--Bi\textsubscript{2}O\textsubscript{3} glasses (TeBaBi) and (b) compositional evolution of magneto-optic anomaly scalling parameter,$\gamma$. }
\end{figure}

\subsection{Magneto-optical properties, Verdet constant}
Magneto-optical properties of prepared glasses were investigated through the estimation of spectral and magnetic field dependency of Faraday rotation defined as $\theta_F = VBl$, where $\theta_F$ represents the angle of rotation of linearly polarized light after passing through the sample of thickness $l$, under an applied magnetic field ($B$) at the direction of light propagation. The obtained spectral and field dependencies of Verdet constant ($V$) for TeBaBi glasses of a known thickness are shown in Fig. \ref{fig:mo}. The experimental values of Verdet constant measured at 632 nm are listed in Table \ref{tab:tebabi_magnetoopt} and vary from 26.7 to 45.3 rad T\textsuperscript{$-$1} m\textsuperscript{$-$1} (0.092$-$0.156 min Oe\textsuperscript{$-$1} cm\textsuperscript{$-$1}). As can be seen from Fig.\ref{fig:tebabi_tern_rho_n_v}, the Verdet constant exhibit practically similar compositional dependence as was observed for refractive index, when MO effect amplitude increases with higher TeO\textsubscript{2} and Bi\textsubscript{2}O\textsubscript{3} content. The highest Verdet constant, $V$\textsubscript{632} = 45.3 radT\textsuperscript{$-$1} m\textsuperscript{$-$1}, was observed for composition Te80Ba10Bi10, which is significantly higher value compared to pure TeO\textsubscript{2} glass (36.3 radT\textsuperscript{$-$1} m\textsuperscript{$-$1})\cite{hrabovsky_pureTeO2} or some previously reported ternary tellurite glasses, such as TeO\textsubscript{2}$-$ZnO$-$BaO (22.0$-$32.8 radT\textsuperscript{$-$1}m\textsuperscript{$-$1})\cite{HRABOVSKY2024_tzboptika}, TeO\textsubscript{2}$-$ZnO$-$La\textsubscript{2}O\textsubscript{3} (29 radT\textsuperscript{$-$1}m\textsuperscript{$-$1})\cite{CHEN2015_TZL_MO} or TeO\textsubscript{2}$-$ZnO$-$Na\textsubscript{2}O (27.1/28.1 radT\textsuperscript{$-$1}m\textsuperscript{$-$1})\cite{CHEN2014_TZN_MO,Chen2015_TZN_MO}. Observed value is also more than one order of magnitude larger than for fused SiO\textsubscript{2} at almost similar used wavelength ($V\textsubscript{633}$ = 3.7 rad T\textsuperscript{$-$1} m\textsuperscript{$-$1})\cite{dshore_2011_verdet_sio2}. Obtained values are then higher than those of other light flint glasses but lower than those of crystals or paramagnetic glasses, which typically contain rare-earth elements that enhance the magneto-optical response\cite{HRABOVSKY2024_tzboptika,RUAN2005,CHEN2015_mo,liu_MO}. Tellurite glasses lacking rare-earth ions are then classified as diamagnetic glasses\cite{HRABOVSKY2024_tzboptika,Chen2015_TZN_MO}, exhibiting a positive rotation angle $\theta$ when subjected to an external magnetic field, $B$. Therefore, the material’s magneto-optical response, represented by Faraday rotation, can be approximated in transparent region using the Sellmeier model (Eq.\ref{eq:sell_mo}), with two fitting parameters $A$\textsubscript{M} and $B$\textsubscript{M}. The experimental data were fitted to the Sellmeier model across the full spectral region where Faraday rotation was measured, excluding the 1200 to 1600 nm range. This exclusion accounts for minor measurement inconsistencies caused by the experimental setup and optics. This led to a more accurate and consistent data fit. Obtained Sellmeier parameters with denoted measurement uncertainties are listed in Table \ref{tab:tebabi_magnetoopt}. 

\begin{equation}
\label{eq:sell_mo}
V = \frac{ A\textsubscript{M}E\textsuperscript{2}}{(B\textsubscript{M}\textsuperscript{2}-E\textsuperscript{2})\textsuperscript{2}}
\end{equation}

\begin{table}[]
\centering
    \caption{Obtained magneto-optical parameters of TeO\textsubscript{2}$-$BaO$-$Bi\textsubscript{2}O\textsubscript{3}  (TeBaBi) glasses: experimental Verdet constant ($V$) at 632 nm and calculated Verdet ($V$\textsubscript{cal}) constant at 632 and 1550 nm via Eq.\ref{eq:becq}, magneto-optic anomaly ($\gamma$) and Sellmeier parameters ($A\textsubscript{M}, B\textsubscript{M}$) for spectral description of Verdet constant using Eq.\ref{eq:sell_mo}. The estimated Verdet constant error range is $\pm$0.2 rad T\textsuperscript{$-$1} m\textsuperscript{$-$1} and shown numbers in parenthesis represent the tolerance on the last digit.  }
\label{tab:tebabi_magnetoopt}
\fontsize{8pt}{8pt}\selectfont
\begin{tabular}{ccccccc}
\hline
\multirow{2}{*}{SID} & $V$\textsubscript{632} & $V$\textsubscript{632,cal} & $V$\textsubscript{1550,cal} & $\gamma$  & $A$\textsubscript{M} & BM   \\
                     & \multicolumn{3}{c}{(radT\textsuperscript{$-$1}m\textsuperscript{$-$1})}    &       &   (radT\textsuperscript{$-$1}m\textsuperscript{$-$1}eV\textsuperscript{2})     &   (eV)   \\ \hline
Te60Ba25Bi15                    & 39.6 & 39.3     & 5.7   & 0.743 & 14.0(1)$\times$10\textsuperscript{3}   & 6.41 \\
Te65Ba20Bi15                    & 41.0 & 41.9     & 6.2   & 0.905 & 14.1(2)$\times$10\textsuperscript{3}   & 6.35 \\ \hdashline
Te55Ba35Bi10                    & 33.4 & 33.6     & 4.9   & 0.738 & 14.3(3)$\times$10\textsuperscript{3}   & 6.68 \\
Te60Ba30Bi10                    & 32.9 & 32.7     & 4.8   & 0.743 & 14.7(1)$\times$10\textsuperscript{3}   & 6.76 \\
Te70Ba20Bi10                    & 37.2 & 37.7     & 5.5   & 0.753 & 16.2(1)$\times$10\textsuperscript{3}   & 6.69 \\
Te80Ba10Bi10                    & 45.3 & 46.2     & 6.7   & 0.920 & 20.9(1)$\times$10\textsuperscript{3}   & 6.77 \\ \hdashline
Te60Ba35Bi5                    & 26.7 & 26.7     & 4.2   & 0.711 & 19.0(1)$\times$10\textsuperscript{3}   & 7.45 \\
Te75Ba20Bi5                     & 38.5 & 39.1     & 5.8   & 0.920 & 22.9(2)$\times$10\textsuperscript{3}   & 7.18 \\
Te85Ba10Bi5                     & 37.9 & 38.2     & 5.6   & 0.852 & 24.0(2)$\times$10\textsuperscript{3}   & 7.29 \\ \hline
Te80Zn10Ba10\cite{HRABOVSKY2024_tzboptika}                    & 31.4 & 30.5     & 4.4   & 0.774 & 9.2(2)$\times$10\textsuperscript{3}   & 6.11 \\ 
Pure TeO\textsubscript{2} (Pt)\cite{hrabovsky_pureTeO2}                    & 36.3 & 32.4     & 4.6  & 0.770 & 25.6(1)$\times$10\textsuperscript{3}   & 7.61 \\ \hline
\end{tabular}
\end{table}

\begin{equation}
\label{eq:becq}
V\textsubscript{D} = \frac{ \gamma e}{2m\textsubscript{e}c} \lambda \frac{dn}{d\lambda}
\end{equation} 

Using the diamagnetic nature of prepared glasses, their MO response may be modelled according to the Becquerel theory of electromagnetism using the Eq.\ref{eq:becq}, where $e$ and $m$\textsubscript{e} represents the unit electron charge and electron mass, $\lambda$ is the light wavelength, c is the speed of light and $\gamma$ is the magneto-optic anomaly. This magneto-optic anomaly multiplicative factor then quantifies the level of agreement between experimental and calculated values of Verdet constant in diamagnetic materials. The Eq.\ref{eq:becq} then presented the connection between the spectral dispersion of optical refractive index, d$n$/d$\lambda$, and magneto-optical effect which is described via Verdet constant. As a result, higher values of the Verdet constant, both theoretical and experimental, are naturally found near the optical band-gap-energy of the material, where the onset of optical absorption begins to have a significant impact. Therefore, to facilitate the description of the dispersion of the refractive index in the transparent region, the Sellmeier model (Eq.\ref{eq:selm_optika_1term}) was used, from which the dispersion term was later derived. A detailed description of the procedure can be found in Ref.\cite{HRABOVSKY2024_tzboptika}. The theoretical  $V$\textsubscript{D}($\lambda$) values calculated at the selected wavelengths were matched to the experimental data $V$($\lambda$) using the least squares method, with the gamma parameter allowed to vary freely. Obtained values of magneto-optic anomaly and calculated Verdet constants, $V$\textsubscript{cal, $\lambda$} ($\lambda$ = 632 nm, 1550 nm), are listed in Table\ref{tab:tebabi_magnetoopt}, with an estimated mean squarred error $R^2>$0.996.

The models based on Eq.\ref{eq:sell_mo} and Eq.\ref{eq:becq} reliably predict the Verdet constant even beyond the experimentally measured range, as long as the extrapolation remains within the material’s transparent spectral region. This allows to estimate the compositional dependence of Verdet constant value at wavelength of $\lambda =$1550 nm, $V$\textsubscript{1550}$\approx$4.2$-$6.7 radT\textsuperscript{$-$1}m\textsuperscript{$-$1}, which is commonly used in tellecommunication devices. Note, that obtained values using both models are in good agreement within the experimental error uncertainty. The observed  magnitude of Faraday rotation at 1550 nm is then comparable to the pure TeO\textsubscript{2} glass ($V$\textsubscript{1550}$\approx$4.6 radT\textsuperscript{$-$1}m\textsuperscript{$-$1})\cite{hrabovsky_pureTeO2} or other TeO\textsubscript{2}-rich multicomponent glasses ($V$\textsubscript{1550}$\approx$4.7 radT\textsuperscript{$-$1}m\textsuperscript{$-$1})\cite{HRABOVSKY2024_tzboptika}. It is also worth noting that $\gamma$ values for pure TeO\textsubscript{2} are close to those obtained for TZB glasses (0.771$-$0.832)\cite{hrabovsky_pureTeO2,HRABOVSKY2024_tzboptika}. This suggests that the MO response of diamagnetic tellurite glasses can be predicted in transparent spectral region with good reliability using the known spectral dispersion of optical refractive index or can be at least applied for materials with high TeO\textsubscript{2} content.

\section{Conclusion}

The linear and non-linear optical and magneto-optical properties of TeO\textsubscript{2}--BaO–Bi\textsubscript{2}O\textsubscript{3} (TeBaBi) glasses were systematically characterized across the full glass-forming region (TeO$_2$: 55–85 mol.\%, BaO: 10–35 mol.\%, Bi$_2$O$_3$: 5–15 mol.\%). Measurements were performed over a broad spectral range extending from approximately 193 to 30,000 nm for optical characterization and from 350 to 2500 nm for magneto-optical analysis.  The measured linear refractive index spans from $n_{1550} \approx$1.922--2.084, while the corresponding non-linear refractive index values, $n_2$, range from $1.63$ to $3.45 \times 10^{-11}$esu. The compositional evolution of the refractive index was successfully described using multilinear regression with respect to both molar and atomic content, yielding correlation coefficients $r \geq 0.993$ and prediction deviation $\Delta n \leq 0.01$, confirming excellent material reproducibility. The optical band gap, evaluated via Tauc–Lorentz model on spectroscopic ellipsometry data, was found to lie between 3.1 and 3.6 eV.

Magneto-optical properties were studied through spectral measurements of the Verdet constant, which ranged from 26.7 to 45.3~rad T$^{-1}$m$^{-1}$ at 632 nm depending on composition. Notably, the highest Verdet constant was obtained for the Te80Ba10Bi10 glass, which also exhibited one of the highest refractive index values. Increasing the TeO\textsubscript{2} and Bi\textsubscript{2}O\textsubscript{3} content led to a systematic rise in both linear and non-linear refractive indices as well as in magneto-optical activity with more pronounced influence of Bi\textsubscript{2}O\textsubscript{3}. The wavelength dispersion of the Verdet constant was analyzed and fitted using a   Becquerel model together with spectral dispersion of linear refractive index, allowing for the extraction of the magneto-optic anomaly parameter $\gamma$. The estimated $\gamma$ values, spanning 0.71–0.92, suggest that the magneto-optical response of diamagnetic tellurite glasses can be reliably predicted from the known spectral dispersion of the optical refractive index.

In comparison to the previously reported TeO$_2$–ZnO–BaO (TZB) glass system prepared and characterized under similar conditions, TeBaBi glasses demonstrate higher Verdet constants and linear/non-linear refractive index values while maintaining similar refractive index tunability. The strong correlation between composition, refractive index, and magneto-optical response makes both TeBaBi and TZB glass systems a versatile and robust candidates for photonic and MO device applications, particularly in the visible to mid-infrared spectral region.

\section*{Acknowledgements}
J.H. gratefully acknowledges the support of the J.W. Fulbright Commission through the Fulbright-Masaryk Scholarship Program. This work was supported by the National Science Foundation under Grant No. DMR--2423992.


\bibliography{mybibfile_checked}

\end{document}